\documentclass[twocolumn]{aastex63}

\usepackage{graphicx}
\usepackage{apjfonts}
\usepackage{amsmath}
\usepackage[raggedright]{subfigure}
\usepackage{color}
\usepackage{hyperref}
\usepackage{acronym}
\usepackage{xspace}

\newcommand{\porb}{P_{\rm orb}}
\newcommand{\tasc}{T_{\rm asc}}
\newcommand{\psr}{PSR\,J1653$-$0158\xspace}
\newcommand{\src}{4FGL\,J1653.6$-$0158\xspace}
\newcommand{\fgbw}{PSR\,J1311$-$3430\xspace}
\def\Fermi{\textit{Fermi}\xspace}
\newcommand{\atlas}{ATLAS\xspace}
\newcommand{\EatH}{\textit{Einstein@Home}\xspace}
\defcitealias{gaia2018}{{\it Gaia}}
\defcitealias{manchester2005}{ATNF}

\shorttitle{Discovery of \psr through Gamma Rays} %%% < 44 characters
\shortauthors{\sc Nieder et al.}

\begin{document}

%% Acronyms
% no plural
\newacro{LAT}[LAT]{Large Area Telescope}
\newacro{3FGL}[3FGL]{\Fermi~\acs{LAT} Third Source Catalog}
\newacro{4FGL}[4FGL]{\Fermi~\acs{LAT} Fourth Source Catalog}
\newacro{ATNF}[\citetalias{manchester2005}]{Australia Telescope National Facility}
\newacro{aLIGO}[aLIGO]{advanced Laser Interferometer Gravitational-wave Observatory}
% normal plural (+s)
\newacro{MSP}[MSP]{millisecond pulsar}
\newacro{GW}[GW]{gravitational wave}
\newacro{IBS}[IBS]{intrabinary shock}

%% Title
\title{Discovery of a Gamma-ray Black Widow Pulsar by GPU-accelerated Einstein@Home}

%% Authorlist
\author[0000-0002-5775-8977]{L.~Nieder}
\affiliation{Max-Planck-Institut f\"ur Gravitationsphysik (Albert-Einstein-Institut), 30167 Hannover, Germany}
\affiliation{Leibniz Universit\"at Hannover, 30167 Hannover, Germany}

\author[0000-0003-4355-3572]{C.~J.~Clark}
\affiliation{Jodrell Bank Centre for Astrophysics, Department of Physics and Astronomy, The University of Manchester, M13 9PL, UK}

\author[0000-0002-5402-3107]{D.~Kandel}
\affiliation{KIPAC/Dept. of Physics, Stanford University, Stanford, CA 94305, USA}

\author[0000-0001-6711-3286]{R.~W.~Romani}
\affiliation{KIPAC/Dept. of Physics, Stanford University, Stanford, CA 94305, USA}

\author[0000-0002-1429-9010]{C.~G.~Bassa}
\affiliation{ASTRON, The Netherlands Institute for Radio Astronomy, Oude Hoogeveensedijk 4, 7991 PD Dwingeloo, The Netherlands}

%alphabetical part I

\author[0000-0003-4285-6256]{B.~Allen}
\affiliation{Max-Planck-Institut f\"ur Gravitationsphysik (Albert-Einstein-Institut), 30167 Hannover, Germany}
\affiliation{Department of Physics, University of Wisconsin--Milwaukee, P.O. Box 413, Milwaukee, WI 53201, USA}
\affiliation{Leibniz Universit\"at Hannover, 30167 Hannover, Germany}

\author[0000-0002-8395-957X]{A.~Ashok}
\affiliation{Max-Planck-Institut f\"ur Gravitationsphysik (Albert-Einstein-Institut), 30167 Hannover, Germany}
\affiliation{Leibniz Universit\"at Hannover, 30167 Hannover, Germany}

\author[0000-0002-1775-9692]{I.~Cognard}
\affiliation{Laboratoire de Physique et Chimie de l'Environnement et de l'Espace, Universit\'e d'Orl\'eans / CNRS, F-45071 Orl\'eans Cedex 02, France}
\affiliation{Station de radioastronomie de Nan\c{c}ay, Observatoire de Paris, CNRS/INSU, F-18330 Nan\c{c}ay, France}

\author[0000-0001-8036-1882]{H.~Fehrmann}
\affiliation{Max-Planck-Institut f\"ur Gravitationsphysik (Albert-Einstein-Institut), 30167 Hannover, Germany}
\affiliation{Leibniz Universit\"at Hannover, 30167 Hannover, Germany}

\author[0000-0003-1307-9435]{P.~Freire}
\affiliation{Max-Planck-Institut f\"ur Radioastronomie, auf dem H\"ugel 69, 53121 Bonn, Germany}

\author[0000-0002-5307-2919]{R.~Karuppusamy}
\affiliation{Max-Planck-Institut f\"ur Radioastronomie, auf dem H\"ugel 69, 53121 Bonn, Germany}

\author[0000-0002-4175-2271]{M.~Kramer}
\affiliation{Max-Planck-Institut f\"ur Radioastronomie, auf dem H\"ugel 69, 53121 Bonn, Germany}
\affiliation{Jodrell Bank Centre for Astrophysics, Department of Physics and Astronomy, The University of Manchester, M13 9PL, UK}

\author[0000-0003-3010-7661]{D.~Li}
\affiliation{National Astronomical Observatories, Chinese Academy of Sciences, Beijing 100101, China}
\affiliation{NAOC-UKZN Computational Astrophysics Centre, University of KwaZulu-Natal, Durban 4000, South Africa}

\author[0000-0002-2332-0459]{B.~Machenschalk}
\affiliation{Max-Planck-Institut f\"ur Gravitationsphysik (Albert-Einstein-Institut), 30167 Hannover, Germany}
\affiliation{Leibniz Universit\"at Hannover, 30167 Hannover, Germany}

\author[0000-0001-7771-2864]{Z.~Pan}
\affiliation{National Astronomical Observatories, Chinese Academy of Sciences, Beijing 100101, China}

\author[0000-0002-1007-5298]{M.~A.~Papa}
\affiliation{Max-Planck-Institut f\"ur Gravitationsphysik (Albert-Einstein-Institut), 30167 Hannover, Germany}
\affiliation{Department of Physics, University of Wisconsin--Milwaukee, P.O. Box 413, Milwaukee, WI 53201, USA}
\affiliation{Leibniz Universit\"at Hannover, 30167 Hannover, Germany}

\author[0000-0001-5799-9714]{S.~M.~Ransom}
\affiliation{National Radio Astronomy Observatory, 520 Edgemont Rd., Charlottesville, VA  USA 22903}

\author[0000-0002-5297-5278]{P.~S.~Ray}
\affiliation{Space Science Division, Naval Research Laboratory, Washington, DC 20375-5352, USA}

\author[0000-0002-2892-8025]{J.~Roy}
\affiliation{National Centre for Radio Astrophysics, Tata Institute of Fundamental Research, Pune 411 007, India}

\author[0000-0002-3386-7159]{P.~Wang}
\affiliation{National Astronomical Observatories, Chinese Academy of Sciences, Beijing 100101, China}

\author[0000-0003-3536-4368]{J.~Wu}
\affiliation{Max-Planck-Institut f\"ur Radioastronomie, auf dem H\"ugel 69, 53121 Bonn, Germany}

%alphabetical part II

\author[0000-0002-1481-8319]{C.~Aulbert}
\affiliation{Max-Planck-Institut f\"ur Gravitationsphysik (Albert-Einstein-Institut), 30167 Hannover, Germany}
\affiliation{Leibniz Universit\"at Hannover, 30167 Hannover, Germany}

\author[0000-0001-8715-9628]{E.~D.~Barr}
\affiliation{Max-Planck-Institut f\"ur Radioastronomie, auf dem H\"ugel 69, 53121 Bonn, Germany}

\author[0000-0002-8524-1537]{B.~Beheshtipour}
\affiliation{Max-Planck-Institut f\"ur Gravitationsphysik (Albert-Einstein-Institut), 30167 Hannover, Germany}
\affiliation{Leibniz Universit\"at Hannover, 30167 Hannover, Germany}

\author[0000-0003-0679-8562]{O.~Behnke}
\affiliation{Max-Planck-Institut f\"ur Gravitationsphysik (Albert-Einstein-Institut), 30167 Hannover, Germany}
\affiliation{Leibniz Universit\"at Hannover, 30167 Hannover, Germany}

\author[0000-0002-6287-6900]{B.~Bhattacharyya}
\affiliation{National Centre for Radio Astrophysics, Tata Institute of Fundamental Research, Pune 411 007, India}

\author[0000-0001-8522-4983]{R.~P.~Breton}
\affiliation{Jodrell Bank Centre for Astrophysics, Department of Physics and Astronomy, The University of Manchester, M13 9PL, UK}

\author[0000-0002-1873-3718]{F.~Camilo}
\affiliation{South African Radio Astronomy Observatory, 2 Fir Street, Black River Park, Observatory 7925, South Africa}

\author[0000-0001-6900-1851]{C.~Choquet}
\affiliation{R\'{e}sidence Le Dauphin\'{e}, rue Jean Bleuzen, Vanves, France}

\author[0000-0003-4236-9642]{V.~S.~Dhillon}
\affiliation{Department of Physics and Astronomy, University of Sheffield, Sheffield S3 7RH, UK}
\affiliation{Instituto de Astrof\'{i}sica de Canarias, E-38205 La Laguna, Tenerife, Spain}

\author[0000-0001-7828-7708]{E.~C.~Ferrara}
\affiliation{NASA Goddard Space Flight Center, Greenbelt, MD 20771, USA}
\affiliation{Department of Astronomy, University of Maryland, College Park, MD 20742, USA}

\author[0000-0002-9049-8716]{L.~Guillemot}
\affiliation{Laboratoire de Physique et Chimie de l'Environnement et de l'Espace, Universit\'e d'Orl\'eans / CNRS, F-45071 Orl\'eans Cedex 02, France}
\affiliation{Station de radioastronomie de Nan\c{c}ay, Observatoire de Paris, CNRS/INSU, F-18330 Nan\c{c}ay, France}

\author[0000-0003-2317-1446]{J.~W.~T.~Hessels}
\affiliation{ASTRON, The Netherlands Institute for Radio Astronomy, Oude Hoogeveensedijk 4, 7991 PD Dwingeloo, The Netherlands}
\affiliation{Anton Pannekoek Institute for Astronomy, University of Amsterdam, Science Park 904, 1098 XH Amsterdam, The Netherlands}

\author[0000-0002-0893-4073]{M.~Kerr}
\affiliation{Space Science Division, Naval Research Laboratory, Washington, DC 20375-5352, USA}

\author{S.~A.~Kwang}
\affiliation{Department of Physics, University of Wisconsin-Milwaukee, P.O. Box 413, Milwaukee, WI 53201, USA}

\author[0000-0002-2498-7589]{T.~R.~Marsh}
\affiliation{Astronomy and Astrophysics Group, Department of Physics, University of Warwick, Coventry CV4 7AL, UK}

\author{M.~B.~Mickaliger}
\affiliation{Jodrell Bank Centre for Astrophysics, Department of Physics and Astronomy, The University of Manchester, M13 9PL, UK}

\author[0000-0002-4795-697X]{Z.~Pleunis}
\affiliation{Department of Physics, McGill University, 3600 rue University, Montr\'{e}al, QC H3A 2T8, Canada}
\affiliation{McGill Space Institute, McGill University, 3550 rue University, Montr\'{e}al, QC H3A 2A7, Canada}

\author[0000-0002-1164-4755]{H.~J.~Pletsch}
\affiliation{Max-Planck-Institut f\"ur Gravitationsphysik (Albert-Einstein-Institut), 30167 Hannover, Germany}

\author[0000-0002-9396-9720]{M.~S.~E.~Roberts}
\affiliation{New York University Abu Dhabi, P.O. Box 129188, Abu Dhabi, UAE}
\affiliation{Eureka Scientific, Inc. 2452 Delmer Street, Suite 100, Oakland, CA 94602-3017, USA}

\author{S.~Sanpa-arsa}
\affiliation{National Astronomical Research Institute of Thailand (Public Organization), 260 Moo 4, T. Donkaew, A. Maerim, Chiang Mai, 50180, Thailand}

\author[0000-0003-1833-5493]{B.~Steltner}
\affiliation{Max-Planck-Institut f\"ur Gravitationsphysik (Albert-Einstein-Institut), 30167 Hannover, Germany}
\affiliation{Leibniz Universit\"at Hannover, 30167 Hannover, Germany}

% corresponding author
\correspondingauthor{L.~Nieder}
\email{lars.nieder@aei.mpg.de}

% dates
\received{2020 September 1}
\revised{2020 September 22}
\accepted{2020 September 25}
\published{2020 October 22}

%%%%%%%%%%%%%%%%%%%%%%%%%%%%%%%%%%%%%%%%%%%
\begin{abstract} %%%<250words
\noindent
We report the discovery of $1.97$\,ms period gamma-ray pulsations from the $75$\,minute orbital-period binary pulsar now named \psr.  The associated \Fermi-\acl{LAT} gamma-ray source \src has long been expected to harbor a binary \acl{MSP}.  Despite the pulsar-like gamma-ray spectrum and candidate optical/X-ray associations -- whose periodic brightness modulations suggested an orbit -- no radio pulsations had been found in many searches.  The pulsar was discovered by directly searching the gamma-ray data using the GPU-accelerated \EatH distributed volunteer computing
system.  The multi-dimensional parameter space was bounded by positional and orbital constraints obtained from the optical counterpart.  More sensitive analyses of archival and new radio data using knowledge of the pulsar timing solution yield very stringent upper limits on radio emission.  Any radio emission is thus either exceptionally weak, or eclipsed for a large fraction of the time.  The pulsar has one of the three lowest inferred surface magnetic-field strengths of any known pulsar with $B_{\rm surf} \approx 4 \times 10^{7}$\,G.  The resulting mass function, combined with models of the companion star's optical light curve and spectra, suggests a pulsar mass $\gtrsim 2\,M_{\odot}$.  The companion is light-weight with mass $\sim 0.01\,M_{\odot}$, and the orbital period is the shortest known for any rotation-powered binary pulsar.  This discovery demonstrates the \Fermi-\acl{LAT}'s potential to discover extreme pulsars that would otherwise remain undetected.
\end{abstract}


%%%%%%%%%%%%%%%%%%%%%%%%%%%%%%%%%%%%%%%%%%%
\section{Introduction}\label{s:introduction}

The \Fermi \ac{LAT} source \src is a bright gamma-ray source, and the brightest remaining unassociated source \citep{parkinson2016}.  It was first seen by the Energetic Gamma Ray Experiment Telescope \citep[EGRET;][]{hartmann1999}, and was also listed in the \ac{LAT} Bright Gamma-ray source list \citep{0fgl} more than a decade ago.  While pulsars were discovered in several other sources from this list \citep[see, e.g.,][]{ransom2011}, the origin of \src remained unidentified.  The detection of a variable X-ray and optical candidate counterpart with $75$\,min period consistent with the gamma-ray position of \src provided strong evidence of it being a binary gamma-ray pulsar \citep{kong2014,romani2014}.

To identify the neutron star in \src, we carried out a binary-pulsar search of the gamma rays, using the powerful GPU-accelerated distributed volunteer computing system \EatH.  Such searches are very computationally demanding, and would take decades to centuries on a single computer while still taking weeks or months on \EatH.  Thus, the search methods are specifically designed to ensure efficiency \citep{nieder2020}.  One key element is the use of constraints derived from optical observations.  The companion's pulsar-facing side is heated by the pulsar wind, leading to a periodically varying optical light curve.  This permits the orbital period $P_{\rm orb}$ and other orbital parameters to be tightly constrained (for a feasible search the uncertainty $\Delta P_{\rm orb}$ needs to be less than a few milliseconds).  In addition, because the sky position of the optical source is typically known to high precision (sub-milliarcsecond level), a search over position parameters is not needed.

Here we present the discovery and analysis of gamma-ray pulsations from \psr in \src.  The pulsar is spinning very rapidly, at a rotational frequency of $508$\,Hz.  The inferred surface magnetic-field strength is one of the lowest of all known pulsars.  The discovery also confirms the $75$\,min orbital period.  This very short orbital period raises interesting questions about the evolutionary path which created the system.

This Letter is organized as follows. In Section~\ref{s:gamma}, we describe the gamma-ray search, detection, and analysis within \ac{LAT} data.  The optical analysis of the pulsar's companion, radio pulsation searches, and a continuous gravitational-wave follow-up search are presented in Section~\ref{s:multiwavelength}.  We discuss the results and conclude in Section~\ref{s:conclusion}.

%%%%%%%%%%%%%%%%%%%%%%%%%%%%%%%%%%%%%%%%%%%
\section{Gamma-ray pulsations}\label{s:gamma}

%%%%%%%%%%%%%%%%%%%%%%%%%%%%%%%%%%%%%%%%%%%	
\subsection{Data preparation}\label{ss:datprep}

We searched for gamma-ray pulsations in the arrival times of photons observed by the \textit{Fermi} LAT \citep{atwood2009} between 2008 August 3 and 2018 April 16 (MJDs $54{,}681$ and $58{,}224$). We included \texttt{SOURCE}-class photons according to the \texttt{P8R2\_SOURCE\_V6} \citep{Pass8} instrument response functions (IRFs)\footnote{See \url{https://fermi.gsfc.nasa.gov/ssc/data/analysis/LAT_essentials.html}}, with reconstructed incidence angles within a $5\degr$ region of interest (RoI) around the putative pulsar position, energies above $100$\,MeV, and zenith angles below $90\degr$.  Here, we used the presumptive companion's position as reported in the {\it Gaia} DR2 Catalog \citep[hereafter \citetalias{gaia2018} catalog;][]{gaia2018}.  The celestial parameters (J2000.0) are $\alpha = 16^{\rm h}53^{\rm m}38\fs05381(5)$ and $\delta = -01\arcdeg58\arcmin36\farcs8930(5)$, with $1\sigma$ uncertainties on the last digits reported in parentheses.

Using the photon incidence angles and energies, we constructed a probability or weight for each photon, $w_j \in [0,1]$, where $j$ labels the photon: $w_j$ is the probability that the $j$th photon originated from the posited source, as opposed to a fore- or background source. These weights were computed by \texttt{gtsrcprob}, using the preliminary \textit{Fermi}-LAT 8\,year source catalog\footnote{\url{https://fermi.gsfc.nasa.gov/ssc/data/access/lat/fl8y/}} as a model for the flux within the RoI without performing a full spectral fit. Weighting the contribution of each photon to a detection statistic in this way greatly increases the search sensitivity \citep{Kerr2011}, and the distribution of weights can be used to predict expected signal-to-noise ratios \citep{nieder2020}.

The data set used here consisted of $N = 354{,}009$ photons, collected over a period of $3{,}542$\,days.  The properties of the detection statistics (semicoherent power $S_1$, coherent power $P_1$, and $H$ statistic) depend upon the lowest moments of the weights, which are
\begin{equation*}
	\sum_{j=1}^{N} w_j   \approx 10266 \,, \,\,
	\sum_{j=1}^{N} w_j^2 \approx 2464 \,, \text{ and }
	\sum_{j=1}^{N} w_j^4 \approx 931 \,.
\end{equation*}
These moments determine the ultimate sensitivity to a particular pulse profile and pulsed fraction, as given in Eq.~(11) in \cite{nieder2020}.

Following the pulsar discovery, we extended this dataset to 2020 February 23 (MJD $58{,}902$), using the latest \texttt{P8R3\_SOURCE\_V2} IRFs \citep{Bruel2018+P305}, a larger maximum zenith angle of $105\degr$, and using the \textit{Fermi}-LAT Fourth Source Catalog \citep[hereafter 4FGL;][]{4fgl} as the RoI model for the photon probability weight computations. 

%%%%%%%%%%%%%%%%%%%%%%%%%%%%%%%%%%%%%%%%%%%	
\subsection{Search}\label{ss:search}

The binary-pulsar search methods are described by \cite{nieder2020}, which are a generalization and extension of the isolated-pulsar search methods from \cite{pletsch2014}.

The searched ranges are guided by the known \ac{MSP} population in the \ac{ATNF} Pulsar Catalogue\footnote{\href{http://www.atnf.csiro.au/research/pulsar/psrcat}{http://www.atnf.csiro.au/research/pulsar/psrcat}} \citep{manchester2005}.  For the spin frequency, we searched $f \in [0,1500]$\,Hz\footnote{The upper limit has been chosen to be sensitive to pulsars spinning at up to $750$\,Hz, which have two-peaked pulse profiles where the peaks are half a rotation apart  \citep[see also][]{pletsch2014}. Note that the current record spin frequency is $716$\,Hz \citep{hessels2006}.}.  The spin-frequency derivative was expected to be in the range $\dot{f} \in [-10^{-13},0]$\,Hz\,s$^{-1}$.

The sky position of the candidate optical counterpart is constrained to high precision in the  \citetalias{gaia2018} catalog, so no astrometric search is required.  The proper motion measured by \citetalias{gaia2018} for the optical counterpart was ignored for the search.

%%%%%%%%%%%%%%%%%%%%%%%%%%%%%%%%%%%%%%%%%%%
\subsubsection{Orbital Constraints from Optical Observations}\label{sss:optical}

The orbital-period estimate of \cite{romani2014} was derived from Southern Astrophysical Research (SOAR), WIYN, and Catalina Sky Survey (CSS) observations.  These were augmented by new 350s SOAR Goodman High Throughput
Spectrograph (GHTS) $g^\prime$, $r^\prime$, $i^\prime$ exposures (63 $g^\prime$, 75 $r^\prime$, 42 $i^\prime$) from MJD $56{,}514.074$ -- $56{,}516.184$, and with the $300$\,s $g^\prime$, $r^\prime$, and $i^\prime$ exposures obtained by \citet{kong2014} using the Wide Field camera (WFC) on the 2.5m Isaac Newton Telescope (INT) on La Palma.  For these two data sets, the scatter about the light-curve trends was appreciably larger than the very small statistical errors; we thus add $0.03$\,mag in quadrature to account for unmodeled fast variability and/or photometry systematics.  To further refine the orbital-period uncertainty, we obtained additional observations in $u^\prime$, $g^\prime$, and $i^\prime$ using the high-speed multi-band imager ULTRACAM \citep{ULTRACAM} on the 4.2m William Herschel Telescope (WHT) on two nights (MJDs $57{,}170$ and $57{,}195$), covering six and three orbits of the binary system, respectively, with a series of 20$\,$s exposures. Conditions were very poor on the first night with seeing $> 5$\,arcsec, particularly at the beginning of the observation. We therefore only used the second night's data for the optical light-curve modeling in Section~\ref{ss:optical}, adding the latter half of the first night's observations for orbital-period estimation. Finally, we obtained further INT+WFC exposures (23 $g^\prime$, 151 $r^\prime$, 45 $i^\prime$) on MJD $57{,}988$ -- $57{,}991$. The $g^\prime$, $r^\prime$, $i^\prime$ filter fluxes were referenced to in-field PanSTARRS catalog sources, and then converted to the Sloan Digital Sky Survey (SDSS) scale. The $u^\prime$ photometry was calibrated against an SDSS standard star observed on MJD $57{,}170$. We estimate $\sim 0.05$\,mag systematic uncertainties in $g^\prime$, $r^\prime$, and $i^\prime$, with uncertainties as large as $\sim 0.1$\,mag in $u^\prime$.

We constrained the orbital period using the multi-band Lomb-Scargle periodogram method \citep[][excluding the $u^\prime$ ULTRACAM data, as the modulation has very low signal-to-noise ratio in this band]{VanderPlas2015+MBLSP}. To infer reasonable statistical uncertainties, we fit for and removed constant magnitude offsets, consistent with our estimated calibration uncertainties, between each night's observations in each band, and additionally rescaled the magnitude uncertainties to obtain a reduced chi-square of unity. This constrained the orbital period to  $\porb = 0.0519447518 \pm 6.0 \times 10^{-9}$\,days, where the quoted uncertainty is the $1\sigma$ statistical uncertainty. For the pulsation search, we chose to search the $3\sigma$ range around this value.

In \citet{romani2014}, the time of the pulsar's ascending node, $T_{\rm asc}$, was estimated from the photometric light curve. However, the optical maximum is distinctly asymmetric (see Section~\ref{ss:optical}), which can bias orbital phase estimates. We therefore used the spectroscopic radial-velocity measurements from \citet{romani2014}, folded at the orbital period obtained above, and fit the phase of a sinusoidal radial-velocity curve, finding $\tasc = {\rm MJD}\,56{,}513.47981 \pm 2.1 \times 10^{-4}$. However, as radial velocities may still be slightly biased by asymmetric heating, we elected to search a wide range around this value, corresponding to $\pm 8\sigma$.

For the projected semimajor-axis parameter $x = a_1 \sin i / c$, we decided to start searching $x \in [0,0.1]$\,s, with the intention to go to larger values in the case of no detection.  For a pulsar mass of $1.6\,M_{\odot}$, this would cover the companion mass range up to $0.2\,M_{\odot}$ and would include companion masses of all known ``black-widow'' systems as well as some of the lower-mass ``redback'' systems \citep{roberts2013,strader2019}.  Here, $a_1$ is the pulsar's semimajor axis, $i$ denotes the inclination angle, and $c$ is the speed of light.  As described in \cite{nieder2020}, we expected $x \in [0,0.2]$\,s based on the companion's velocity amplitude reported by \cite{romani2014} and the masses expected for ``spider'' companions, i.e. black-widow or redback companions.

%%%%%%%%%%%%%%%%%%%%%%%%%%%%%%%%%%%%%%%%%%%
\subsubsection{Search grids}\label{sss:grid}

To cover the relevant orbital-parameter space in $\{x,\porb,\tasc\}$, we use \textit{optimized grids} \citep{fehrmann2014}.  These grids use as few points as possible still ensuring that a signal within the relevant space should be detected.  Furthermore, they are able to cover the orbital-parameter space efficiently even though the required density depends on one of the orbital parameters, $x$.

Key to building an optimized grid is to know how the signal-to-noise ratio drops due to offsets from the true pulsar parameters.  This is estimated using a \textit{distance metric} on the orbital-parameter space \citep{nieder2020}.  In our case, the three-dimensional grid was designed to have a worst-case mismatch $\bar{m} = 0.2$, i.e. not more than $20\%$ of the (semicoherent or coherent) signal power should be lost due to orbital-parameter offsets.  Of most relevance is that $99\%$ of randomly injected orbital-parameter points have a mismatch below $\bar{m} = 0.04$ to the closest grid point.

Due to the $f$-dependency of the required grid-point density, we search $f$ in steps, and build the corresponding orbital grids prior to the start of the search on the computing cluster \atlas in Hannover \citep{aulbert2008}.

%%%%%%%%%%%%%%%%%%%%%%%%%%%%%%%%%%%%%%%%%%%
\subsubsection{Einstein@Home}\label{sss:eath}

Searching the $5$-dimensional parameter space $\{f, \dot{f}, x,\porb,\tasc\}$ is a huge computational task with over $10^{17}$ trials.  Thus, the first (computing-intensive) search stages were performed on \EatH, a distributed volunteer computing system \citep{allen2013}.  As done for radio pulsar searches previously, the search code utilizes the approximately $10{,}000$ GPUs active on \EatH for a computing speed-up of $\sim 10$, comparing the runtimes on CPUs and GPUs.

The parameter space is divided into more than one million regions.  Searching one of these is called a ``work unit''.  These work units are sent to computers participating in \EatH, and are searched when the computer is otherwise idle.  Depending on the system, searching a work unit takes between half an hour and up to a few hours of computational time.  In total, the search would have taken more than $50$ years on a single computer, but using \EatH it took less than two weeks.

%%%%%%%%%%%%%%%%%%%%%%%%%%%%%%%%%%%%%%%%%%%
\subsubsection{Gamma-ray detection}\label{sss:detection}

The search process involves multiple stages in which semicoherent statistics are constructed, and the most significant candidates are passed on to fully coherent follow-up stages \citep[for full details of the search pipeline and signal-to-noise ratio definitions, see][]{nieder2020}.  In the last semicoherent stage, a candidate found at a frequency of $1016$\,Hz had signal-to-noise ratio $S_1 = 8.6$, which we now associate with \psr.  This was not the strongest candidate or far above the background of noise, but was among the ten most significant candidates in its work unit, and therefore passed on to the coherent stage.  In the coherent stage, it was very significant, with a signal-to-noise ratio $P_1/2 = 94$.

The search follow-ups confirmed significant pulsations with period $P \approx 1.97$\,ms (or $f \approx 508$\,Hz), while the actual search revealed an alias at twice the pulsar frequency.  This may be because the signal has significant power in the second harmonic.

Note that the signal was found outside the $3\sigma$ range in $\tasc$ from the constraints reported in this work, and outside the $3\sigma$ range given by \cite{romani2014}.  This can be caused by asymmetric heating (see Section~\ref{sss:optical}).

%%%%%%%%%%%%%%%%%%%%%%%%%%%%%%%%%%%%%%%%%%%	
\subsection{Timing}\label{ss:timing}

The parameters used in the phase model to describe the pulsar's rotation are measured in a timing analysis.  We use the timing methods as explained in \cite{clark2017}, which are an extension of the methods by \cite{kerr2015b}.  The basic principle is that the parameter space around the discovery parameters is explored using a Monte Carlo sampling algorithm with a template pulse profile.

To marginalize over the pulse-profile template, we vary the template parameters as described in \cite{nieder2019}.  In the case of \psr, we used a template consisting of two symmetrical, wrapped Gaussian peaks.  We used constraints on the peaks' full-width at half maximum (FWHM), such that the peaks must be broader than $5\%$ of a rotation, and narrower than half a rotation.

\begin{figure}
\centering
\includegraphics[width=\columnwidth]{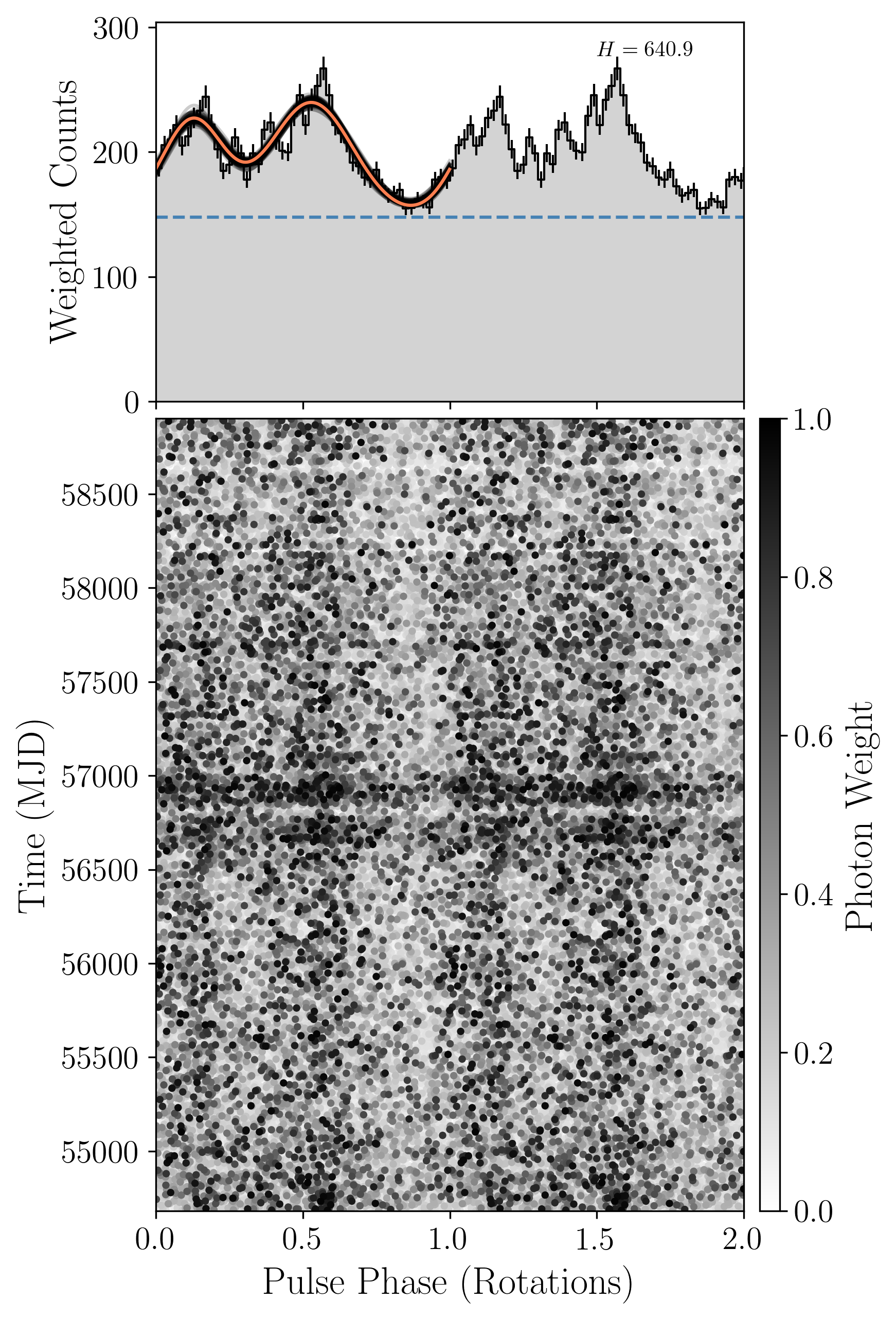}
\caption{\label{f:tvsph} Integrated pulse-profile and phase-time diagram of \psr, showing two identical rotations.  Top: The histogram shows the weighted counts for $50$ bins.  The orange curve indicates the pulse-profile template with the highest signal power, and the transparent black curves represent $100$ templates randomly selected from the Monte Carlo samples after the chain stabilized, to indicate the uncertainty on the profile.  The dashed blue line denotes the source background.  Bottom: Each point represents the pulsar's rotational phase at emission of a photon, with the intensity indicating the photon's probability weight.  Note that \psr received more exposure between MJDs $56{,}600$ and $57{,}000$ when the \ac{LAT} pointed more often toward the Galactic center.}
\end{figure}

Our timing solution over $11$ years of \ac{LAT} data is shown in Table~\ref{t:timing}.  The folded gamma-ray data and the pulse profile are portrayed in Fig.~\ref{f:tvsph}.

\begin{deluxetable}{lc}
\tablewidth{0.99\columnwidth}
\tablecaption{\label{t:timing} Timing solution for \psr. }
\tablecolumns{2}
\tablehead{ Parameter & Value }
\startdata
Range of observational data (MJD)
& $54682$ -- $58902$ \\
Reference epoch (MJD)
& $56100.0$ \\
\cutinhead{Celestial parameters from \citetalias{gaia2018} catalog}
R.A., $\alpha$ (J2000.0)
& $16^{\rm h}53^{\rm m}38\fs05381(5)$ \\
Decl., $\delta$ (J2000.0)
& $-01\arcdeg58\arcmin36\farcs8930(5)$ \\
Positional epoch (MJD)
& $57205.875$ \\
Proper motion in R.A., $\mu_{\alpha} \cos\delta$ (mas yr$^{-1}$)
& $-19.62 \pm 1.86$\\
Proper motion in Decl., $\mu_{\delta}$ (mas yr$^{-1}$)
& $-3.74 \pm 1.12$\\
Parallax\tablenotemark{a}, $\varpi$ (mas)
& $1.88 \pm 1.01$ \\
\cutinhead{Timing parameters}
Spin frequency, $f$ (Hz)
& $508.21219457426(6)$ \\
Spin-frequency derivative, $\dot{f}$ (Hz\,s$^{-1}$)
& $-6.204(8) \times 10^{-16}$ \\
Spin period, $P$ (ms)
& $1.9676820247057(2)$ \\
Spin-period derivative, $\dot{P}$ (s\,s$^{-1}$)
& $2.402(3) \times 10^{-21}$ \\
Proj. semimajor axis, $x$ (s)
& $0.01071(1)$ \\
Orbital period, $\porb$ (days)
& $0.0519447575(4)$ \\
Epoch of ascending node, $\tasc$ (MJD)
& $56513.479171(8)$ \\
\cutinhead{Derived parameters for distance $d = 840$\,pc}
Shklovskii spin-down, $\dot{P}_{\rm Shk}$ (s\,s$^{-1}$)
& $1.6 \times 10^{-21}$ \\
Galactic acceleration spin-down, $\dot{P}_{\rm Gal}$ (s\,s$^{-1}$)
& $-4.8 \times 10^{-23}$ \\
Spin-down power, $\dot{E}$ (erg s$^{-1}$)
& $4.4 \times 10^{33}$ \\
Surface $B$-field, $B_{\rm surf}$ (G)
& $4.1 \times 10^{7}$ \\
Light-cylinder $B$-field, $B_{\rm LC}$ (G)
& $5.0 \times 10^{4}$ \\
Characteristic age, $\tau_{\rm c}$ (Gyr)
& $37$ \\
Gamma-ray luminosity\tablenotemark{b}, $L_{\gamma}$ (erg s$^{-1}$)
& $2.9 \times 10^{33}$ \\
Gamma-ray efficiency, $n_{\gamma} = L_{\gamma} / \dot{E}$
& $0.66$ \\
\enddata
\tablecomments{The JPL DE405 solar system ephemeris has been used, and times refer to TDB.}
\tablenotetext{a}{Corresponds to a model-independent distance $d = 533_{-187}^{+625}$\,pc, but for the derived parameters the consistent distance $d = 840_{-40}^{+40}$\,pc derived from optical modeling is used (see Table~\ref{t:fit_J1653}).}
\tablenotetext{b}{Taken from 4FGL Source Catalog \citep{4fgl}.}
\end{deluxetable}

The observed spin-down $\dot{P}$ is one of the lowest of all known pulsars.  To estimate the intrinsic $\dot{P}$ we account for the Shklovskii effect \citep{shklovskii1970}, and the Galactic acceleration \citep[see, e.g.,][]{damour1991}.  The results are summarized in Table~\ref{t:timing}.  The observed contribution due to the difference in Galactic acceleration of the Sun and the pulsar is computed with $R_{\rm Sun} = 8.21$\,kpc, $z_{\rm Sun} = 14$\,pc, and the Galactic potential model \texttt{PJM17\_best.Tpot} \citep{mcmillan2017}, as implemented in their code\footnote{\href{https://github.com/PaulMcMillan-Astro/GalPot}{https://github.com/PaulMcMillan-Astro/GalPot}}.  For \psr, we used $R_{\rm J1653} = 7.48$\,kpc, and $z_{\rm J1653} = 367$\,pc, assuming $d = 840$\,pc (see Table~\ref{t:fit_J1653}).  The contributions parallel and perpendicular to the Galactic disk nearly cancel each other, so that the choice of the potential and its relevant parameters have a seemingly large effect on the actual small value of $\dot{P}_{\rm Gal}$, and can even change the sign.  However, the overall kinematic contribution to the observed $\dot{P}$ is dominated by the Shklovskii term, and its uncertainty by the uncertainty in the distance estimate.  The estimated intrinsic spin-down is $\dot{P}_{\rm int} = 8.5 \times 10^{-22}$\,s\,s$^{-1}$ for distance $d = 840$\,pc.

%%%%%%%%%%%%%%%%%%%%%%%%%%%%%%%%%%%%%%%%%%%
\section{Multiwavelength and Multimessenger}\label{s:multiwavelength}

\subsection{Optical Light-curve Modeling and System Masses} \label{ss:optical}

\begin{figure}
	\centering
	\includegraphics[width=\columnwidth]{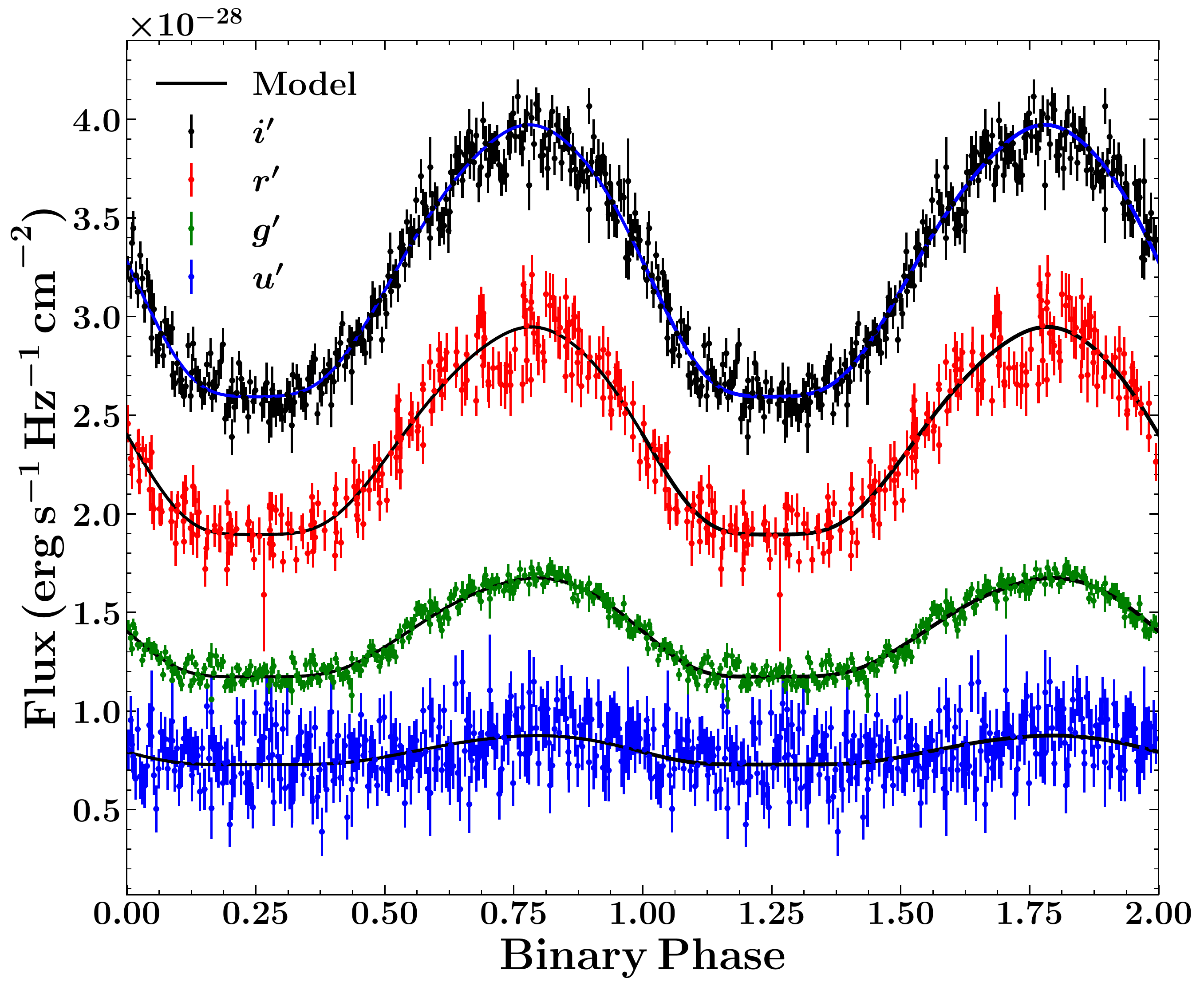}
	\caption{\label{f:lightcurve} $u^\prime$, $g^\prime$, $r^\prime$, and $i^\prime$ light curves for \psr, with the best-fit model curves.  Note the flat minima and decreasing modulation for bluer colors, a consequence of the hard spectrum veiling flux.  Two identical cycles are shown for clarity.}
\end{figure}

By modeling the optical light curves and radial velocities we can constrain the binary mass and distance and the system viewing angle.  Comparing the individual filters between nights suggest small $\delta m \approx 0.05$ shifts in zero-points, consistent with the systematic estimates above.  Correcting to match the individual filters, we then re-binned the light curve, placing the photometry on a regular grid
with points spaced by  $\delta \phi = 0.004$, using the Python package \texttt{Lightkurve}; after excision of a few obviously discrepant points, we retain 248 $u^\prime$, 239 $g^\prime$, 220 $r^\prime$ and 245 $i^\prime$ points for light-curve fitting (Fig.~\ref{f:lightcurve}).  This fitting is done with a version of the \texttt{Icarus} code of \citet{breton2013} modified to include the effect of hot spots on the companion surface, likely generated by precipitation of particles from the \ac{IBS} to companion magnetic poles \citep{Sanchez2017+Bduct}.  All parameter values and errors are determined by Markov Chain Monte Carlo (MCMC) modeling.

The very shallow modulation of these light curves might normally be interpreted as indicating a small inclination $i$.  However given the large companion radial-velocity amplitude $K=666.9\pm7.5\,{\rm km\,s^{-1}}$, implying a mass function $f(M)=1.60\pm0.05\,M_\odot$, measured by \citet{romani2014}, a small inclination would give an unphysical, large neutron star mass.  As noted in that paper, the light curves and spectra show that a strong blue non-thermal veiling flux dominates at orbital minimum.  With increasingly shallow modulation for the bluer colors, this is also evident in the present photometry.  Thus, the minimal model for this pulsar must include a non-thermal veiling flux.  Although this is likely associated with the \ac{IBS}, we model it here as a simple power law with form $f_\nu = f_A (\nu/10^{14}\,{\rm Hz})^{-p}$.  This flux is nearly constant through the orbit, although there are hints of phase structure, e.g. in $r^\prime$ and $i^\prime$ at $\phi_B=0.72$ (see Fig.~\ref{f:lightcurve}).  Any model without such a power-law component is completely unacceptable.  These fits prefer an $A_V$ slightly higher than, but consistent with, the maximum in this direction \citep[obtained by $\sim 300$\,pc;][]{green2019}\footnote{\url{https://doi.org/10.7910/DVN/2EJ9TX}}.

In Fig.~\ref{f:lightcurve}, one notices that the orbital maximum is slightly delayed from $\phi_B=0.75$, especially in the bluer colors.  Such asymmetric heating is most easily modeled adding a polar hot spot with location $(\theta_c, \phi_c)$ and local temperature increase $A_c$ in a Gaussian pattern of width $\sigma_c$; when we include such a component, the fit improves greatly, with $\Delta \chi^2/{\rm DoF} = -0.34$.  The Akaike information criterion (AIC) comparison of the two models indicates that the model with a hot spot is preferred at the $10^{-18}$ level, despite the extra degrees of freedom. We give the fit parameters for both models in Table~\ref{t:fit_J1653}.  Note that with the fine structure near maximum, the model is not yet fully acceptable ($\chi^2/{\rm DoF} \sim 1.4$).  More detailed models, including direct emission from the \ac{IBS} or possibly the effects of companion global winds \citep{kandel2020}, may be needed to fully model the light curves.  Such modeling would be greatly helped by light curves over an even broader spectral range, with \ac{IBS} effects increasingly dominant in the UV, and low-temperature companion emission better constrained in the IR. With many cycles we could also assess the reality (and stability) of the apparent fine structure and test for hot-spot motion.

\begin{deluxetable}{lcc}[t!!]
	\tablecaption{\label{t:fit_J1653} Light-curve fit results for \psr}
	\tablehead{
		\colhead{Parameters} & \colhead{Veiled} & \colhead{Veiled+HS}}
	\startdata
	Inclination, $i$ (deg) & $79.4^{+5.7}_{-6.8}$ & $72.3^{+5.0}_{-4.9}$ \\
	Filling factor, $f_c$ & $0.97^{+0.02}_{-0.02}$  & $0.88^{+0.03}_{-0.03}$ \\
	Heating luminosity, $L_{\mathrm{P}}$ ($10^{33} \mathrm{erg}\,\mathrm{s}^{-1}$) &$3.33^{+0.39}_{-0.34}$&$3.15^{+0.26}_{-0.27}$ \\
	Night-side temperature, $T_N$ (K) & $3250^{+243}_{-331}$ &$3295^{+227}_{-300}$ \\
	$V$-band extinction, $A_V$ & $1.06^{+0.08}_{-0.10}$ & $1.06^{+0.07}_{-0.09}$\\
	Distance, $d$ (pc) & $830^{+50}_{-50}$ &$840^{+40}_{-40}$ \\
	Veiling flux norm, $f_A$ ($\mu$Jy) &$101.7^{+11.4}_{-11.1}$& $99.9^{+11.7}_{-11.4}$ \\
	Veiling flux index, $p$& $0.50^{+0.05}_{-0.03}$& $0.49^{+0.03}_{-0.03}$ \\
	Spot azimuth, $\theta_c$ (deg) &...&$286.8^{+5.8}_{-6.9}$ \\
	Spot co-latitude, $\phi_c$ (deg) &...&$-50.5^{+9.2}_{-8.4}$ \\
	Gaussian spot width, $\sigma_c$ (deg) &...&$25.2^{+5.0}_{-4.9}$ \\
	Spot temperature increase, $A_c$  &...&$0.66^{+0.21}_{-0.21}$ \\
	Neutron star mass, $M_{\rm NS}$ ($M_\odot$)  &$1.99^{+0.18}_{-0.08}$ & $2.17^{+0.21}_{-0.15}$ \\
	Companion mass, $M_{\rm c}$ ($M_\odot$) &$0.013^{+0.001}_{-0.001}$ & $0.014^{+0.001}_{-0.001}$ \\
	$\chi^2/\mathrm{DoF}$ &$1.72$  & $1.38$
	\enddata
	\tablecomments{Parameters from the best-fit light-curve/radial-velocity models, with and without a surface hot spot, including MCMC errors.}
\end{deluxetable}
% other parameters for the record 
% Zero point offsets(mag)     PL                      PL+HS
%du			0.20 +/- 0.07		0.17 +/- 0.06
%dg		   -0.04 +/-  0.04		-0.03 +/- 0.04
%dr		   -0.04 +/- 0.04		-0.02 +/- 0.04
%di			0.01 +/- 0.05		0.02 +/- 0.05
%
%  f_Corr negligibly sensitive to the MCMC chain parameters.

Our fit distance may be cross-checked with two other quantities.  (1) With the 4FGL energy flux $f_\gamma = 3.5 \times 10^{-11}\,{\rm erg\,cm^{-2}\,s^{-1}}$ between $100$\,MeV and $100$\,GeV, our fit distance gives an isotropic gamma-ray luminosity $L_\gamma = 3 \times 10^{33}\,{\rm erg\,s^{-1}}$, in good agreement with the $L_\gamma \approx (10^{33}\,{\rm erg\,s^{-1}}{\dot E})^{1/2}$ heuristic luminosity law \citep{abdo2013}, as a function of the spin-down power $\dot E$.  This luminosity is consistent with the model for direct radiative heating of the companion.  (2) Our fit distance is also consistent with the model-independent, but lower-accuracy, distance from the \citetalias{gaia2018} parallax.  Thus, the $840$\,pc distance seems reliable, although systematic effects probably dominate over the rather small $\sim 50$\,pc statistical errors.

Armed with the fits, we can estimate the companion masses, correcting the observed radial-velocity amplitude (fit with a K-star template) for the temperature-dependent weighting of the absorption lines across the companion face as in \citet{kandel2020}.  The results indicate substantial mass accretion, as expected for these ultra-short-period systems.  With the preferred Veiled+HS model the mass significantly exceeds $2.0\,M_\odot$, adding to the growing list of spider binaries in this mass range.  Note that the inclination $i$ uncertainty dominates the error in this mass determination.  Broader range photometric studies, with better constraint on the heating pattern, can reduce the $i$ uncertainty.

\subsection{Radio pulsation searches}\label{ss:radio}

The pulsar position has been observed in radio multiple times.  Several searches were performed before the gamma-ray pulsation discovery, and a few very sensitive follow-up searches afterward.  Despite the more than $20$ observations with eight of the most sensitive radio telescopes, no radio pulsations have been found.

The results of the radio searches are given in Table~\ref{t:radio}.  Observations are spread over $11$ years, with observing frequencies ranging from $100$\,MHz up to $5$\,GHz.  All orbital phases have been covered by most of the telescopes.  Since there was no detection, the table also gives upper limits derived from the observations.  For all but LOFAR, the data (both archival and recent) were folded with the gamma-ray-derived ephemeris, and searched only over dispersion measure.

The strictest upper limits on pulsed radio emission are $8$\,$\mu$Jy at $1.4$\,GHz, and $20$\,$\mu$Jy at $4.9$\,GHz.  This is fainter than the threshold of $30$\,$\mu$Jy that \cite{abdo2013} use to define a pulsar to be ``radio-quiet''.  Note, that for the calculation of the limits we included the parts of the orbit where eclipses might be expected for spider pulsars.  Thus, the limit constrains the maximum emission of the system, and not the maximum emission from the pulsar alone.

\begin{deluxetable*}{lcccccc}
\tablewidth{0.99\columnwidth}
\tablecaption{\label{t:radio} Summary of radio searches for \psr}
\tablecolumns{5}
\tablehead{ Telescope & Frequency (MHz) & Data start (UTC) & Data span (s) & Orbital phase & Limit ($\mu$Jy) & Reference / Survey }
\startdata
Effelsberg & $1210$--$1510$ & 2010 May 26, 21:33 &  1920 & $0.88$--$1.31$ & $63$ &  \cite{barr2013} \\
Effelsberg & $1210$--$1510$ & 2014 Aug 26, 20:27 &  4600 & $0.15$--$1.17$ & $41$ &  \\
Effelsberg & $4608$--$5108$ & 2014 Aug 29, 18:52 &  4600 & $0.62$--$1.65$ & $33$ &  \\
Effelsberg & $4608$--$5108$ & 2020 Jun 18, 22:09 & 11820 & $0.85$--$3.48$ & $20$ &  \\
FAST       & $1050$--$1450$ & 2020 Jun 04, 16:30 &  2036 & $0.80$--$1.25$ &  $8$ &  \cite{li2018b} \\
GBT        & $720$--$920$   & 2009 Sep 20, 00:49 &  3200 & $0.93$--$1.65$ & $51$ &  \\
GBT        & $720$--$920$   & 2010 Dec 13, 21:04 &  1300 & $0.91$--$1.20$ & $80$ &  \\
GBT        & $720$--$920$   & 2011 Dec 22, 12:11 &  2400 & $0.74$--$1.27$ & $59$ &  \cite{sanpaarsa2016} \\
GBT        & $305$--$395$   & 2012 Feb 22, 14:31 &  1700 & $0.27$--$0.65$ & $301$ &  \\
GBT        & $1700$--$2300$ & 2014 Nov 18, 14:28 &  1200 & $0.36$--$0.63$ & $43$ &  \\
GBT        & $1700$--$2300$ & 2014 Nov 20, 13:56 &  2400 & $0.44$--$0.98$ & $30$ &  \\
GBT        & $1700$--$2300$ & 2014 Nov 21, 22:38 &  1800 & $0.66$--$1.07$ & $35$ &  \\
GBT        & $720$--$920$   & 2017 Jan 28, 13:20 &  1200 & $0.97$--$1.24$ & $83$ &  \\
GMRT       & $591$--$623$   & 2011 Feb 02, 02:32 &  1800 & $0.94$--$1.34$ &   $730$ & \citeauthor{bhattacharyya2013} \\
GMRT       & $306$--$338$   & 2012 May 15, 22:31 &  1800 & $0.54$--$1.06$ &   $990$ & (\citeyear{bhattacharyya2013}, 2020, in prep.) \\
GMRT       & $306$--$338$   & 2012 Jun 11, 17:49 &  1800 & $0.55$--$0.95$ &   $990$ & " \\
GMRT       & $591$--$623$   & 2014 Aug 19, 13:44 &  1800 & $0.00$--$0.54$ &   $270$ & " \\
GMRT       & $591$--$623$   & 2014 Aug 30, 11:17 &  1800 & $0.80$--$1.38$ &   $270$ & " \\
GMRT       & $591$--$623$   & 2015 Dec 28, 03:55 &  1800 & $0.73$--$1.13$ &   $270$ & " \\
LOFAR      & $110$--$180$   & 2017 Mar 15, 04:18 & $15\times320$ & Full orbit  & $6{,}200$ & \cite{bassa2017b}\\
LOFAR      & $110$--$180$   & 2017 Apr 15, 02:20 & $15\times320$ & Full orbit  & $6{,}200$ & " \\
Lovell     & $1332$--$1732$ & 2019 Mar 15, 01:34 &  5400 & $0.57$--$1.77$ & $82$  &  \\
Lovell     & $1332$--$1732$ & 2019 Mar 16, 02:53 &  5400 & $0.87$--$2.08$ & $82$  &  \\
Lovell     & $1332$--$1732$ & 2019 Mar 17, 01:47 &  5400 & $0.25$--$1.45$ & $82$  &  \\
Nan\c{c}ay & $1230$--$1742$ & 2014 Aug 20, 18:33 &  1850 & $0.12$--$0.53$ & $77$  & \cite{desvignes2013} \\
Parkes     & $1241$--$1497$ & 2016 Nov 05, 06:17 &  3586 & $0.26$--$1.06$ & $178$ & \cite{camilo2016} \\
\enddata
\tablecomments{The columns show the telescope used, the observed frequency range, the start time and data span, the range of orbital phases covered, the resulting limit on a pulsed component, and a reference with relevant details.  The orbital phase is given in orbits, and ranges $>1$ indicate that more than one orbit has been observed.  The considered maximum dispersion measure varies with the observing frequency from ${\rm DM} = 80$\,pc\,cm$^{-3}$ at the lowest frequencies to ${\rm DM} = 350$\,pc\,cm$^{-3}$ at the highest frequencies.  To estimate the limit on the pulsed component, we used Eq.~(6) from \cite{ray2011} assuming a pulse width of $0.25\,P$, and a threshold signal-to-noise ratio S/N$_{\rm min} = 7$.}
\end{deluxetable*}

\subsection{Continuous gravitational waves}\label{ss:CW}

We search for nearly monochromatic, continuous \acp{GW} from \psr, using data from the first\footnote{\href{https://doi.org/10.7935/K57P8W9D}{https://doi.org/10.7935/K57P8W9D}} and second\footnote{\href{https://doi.org/10.7935/CA75-FM95}{https://doi.org/10.7935/CA75-FM95}} observing runs of the Advanced LIGO detectors \citep{O1O2_LIGO_2019}.  We assume that \acp{GW} are emitted at the first and second harmonic of the neutron star's rotational frequency, as would occur if the spin axis is misaligned with the principal axes of the moment of inertia tensor \citep{jones2010,jones2015}.

We employ two different analysis procedures, which yield consistent results.  The first is frequentist, based on the multi-detector maximum-likelihood $\mathcal{F}$-statistic introduced by \cite{cutler2005}.  The second is the Bayesian time-domain method \citep{dupuis2005} as detailed by \cite{pitkin2017}, with triaxial non-aligned priors \citep{pitkin2015}.  Both methods coherently combine data from the two detectors, taking into account their antenna patterns and the \ac{GW} polarization.  The $\mathcal{F}$-statistic search excludes data taken during times when the relevant frequency bands are excessively noisy.

The results are consistent with no \ac{GW} emission.  At twice the rotation frequency, the $\mathcal{F}$-statistic $95\%$ confidence upper limit on the intrinsic \ac{GW} amplitude $h_0$ is $4.4 \times 10^{-26}$.  The $95\%$ credible interval upper limit from the Bayesian analysis on $h_0 = 2 C_{22}$ is $3.0 \times 10^{-26}$.  At the rotation frequency (only checked with the Bayesian method) the $95\%$ confidence upper limit on the amplitude $C_{21}$ is $6.6 \times 10^{-26}$.

Since the dominant \ac{GW} frequency might be mismatched from twice the rotation frequency \citep{abbott2019a}, we performed an $\mathcal{F}$-statistic search in a $\pm 1$\,Hz band around this, with an extended $\dot{f}$-range.  This yields larger upper limits on $h_0$, with mean value of $1.3 \times 10^{-25}$ in $10$\,mHz-wide bands.  Full details are given in the supplementary materials.

Our upper limits on $h_0$ at twice the rotation frequency may also be expressed as upper limits on the ellipticity $\epsilon$ of the pulsar \citep{abbott2019b}.  This is $\epsilon = 3.9\times 10^{-8} \times ({h_0 / {5 \times 10^{-26}}}) \times (10^{45} \textrm{g\,cm}^3 / I_{zz}) \times ({{840~\textrm{pc}} / {d}})$, where $I_{zz}$ is the moment of inertia about the spin axis, and $d$ is the distance.

As is the case for most known pulsars, it is unlikely that our searches would have detected a \ac{GW} signal.  In fact, suppose that all of the rotational kinetic-energy losses associated with the intrinsic spin-down are via \ac{GW} emission.  Then assuming the canonical $I_{zz} = 10^{45} \textrm{g\,cm}^3$, this would imply a ``spin-down'' ellipticity $\epsilon^{\textrm{sd}} = 4.7 \times 10^{-10}$, which is a factor $\sim 80$ below our upper limit.

%%%%%%%%%%%%%%%%%%%%%%%%%%%%%%%%%%%%%%%%%%%
\section{Discussion and Conclusions}\label{s:conclusion}

\psr is the second binary pulsar \citep{pletsch2012} and the fourth \ac{MSP} \citep{clark2018} to be discovered through periodicity searches of gamma rays.  This pulsar is remarkable in many ways.  It is only the second rotationally powered \ac{MSP} from which no radio pulsations have been detected.  It is among the fastest-rotating known pulsars with spin frequency $f = 508$\,Hz.  The $75$\,min orbital period is shorter than for any other known rotation-powered pulsar, with the previous record being \fgbw with a $93$\,min orbit \citep{pletsch2012}.  The inferred surface magnetic field is possibly the weakest, depending on the Shklovskii correction.

The discovery was enabled by constraints on the sky-position and orbital parameters from optical observations, together with efficient search techniques and the large computing power of the distributed volunteer computing system \EatH.  The detection proves that the optically variable candidate counterpart \citep{kong2014,romani2014} is indeed the black-widow-type binary companion to \psr, and it conclusively resolves the nature of the brightest remaining unidentified gamma-ray source, first found more than two decades ago \citep{hartmann1999}.

The distance to \psr and its proper motion are well constrained.  \citetalias{gaia2018} measurements of the parallax, $\varpi = 1.88 \pm 1.01$\,mas, imply a distance $d = 530_{-200}^{+470}$\,pc.  A consistent, but tighter constraint is given by our optical modeling with $d = 840_{-40}^{+40}$\,pc.  The proper motion (see Table~\ref{t:timing}) is also measured with good precision (\citetalias{gaia2018} and our timing are in agreement).

\begin{figure}
	\centering
	\includegraphics[width=\columnwidth]{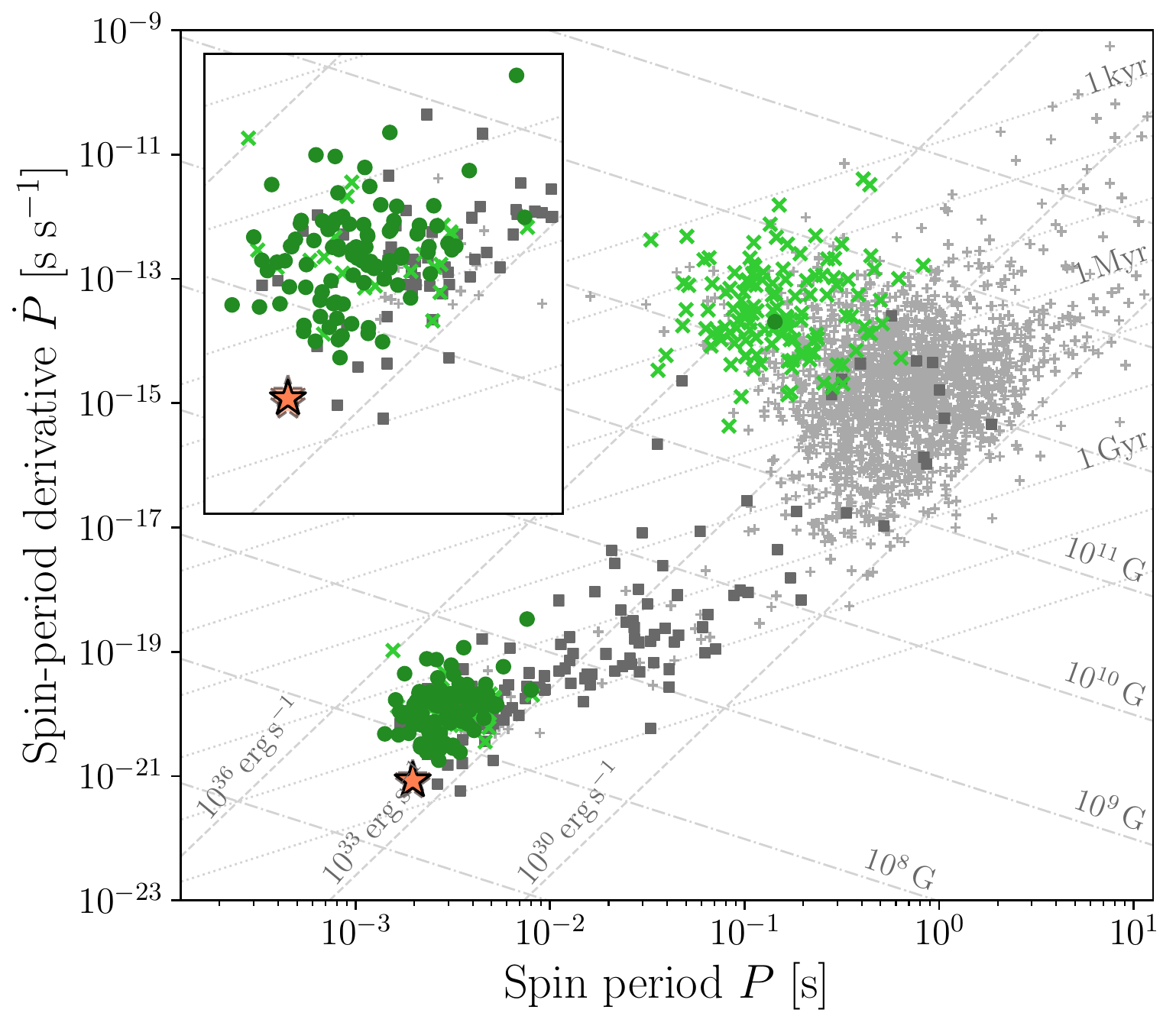}
	\caption{\label{f:ppdot} Newly detected \psr on a $P$--$\dot{P}$ diagram of the known pulsar population outside of globular clusters.  The \ac{MSP} population is shown magnified in the inset.  \ac{LAT} pulsars are marked in green (isolated by a cross and binary by a circle).  Non-\ac{LAT} pulsars in the \ac{ATNF} are marked in gray (isolated by a plus and binary by a square).  The lines show constant surface magnetic-field strength (dashed-dotted), characteristic age (dotted), and spin-down power (dashed).  The spin period and intrinsic spin-period derivative of \psr are marked by the orange star.  The transparent stars indicate the (distance-dependent) maximum and minimum intrinsic spin-period derivatives according to the distance estimated from our optical models.}
\end{figure}

\psr has one of the lowest observed spin-period derivatives of all known pulsars ($\dot{P} = 2.4 \times 10^{-21}\,\textrm{s}\,\textrm{s}^{-1}$).  The intrinsic $\dot{P} = 8.5 \times 10^{-22}\,\textrm{s}\,\textrm{s}^{-1}$ (accounting for Galactic acceleration and Shklovskii effects) is even smaller.  In Fig.~\ref{f:ppdot}, \psr is shown in a $P$-$\dot{P}$ diagram, alongside the known radio and gamma-ray pulsar population outside of globular clusters.

The intrinsic $\dot{P}$ can be used to estimate the pulsar's spin-down power $\dot{E}$, surface magnetic-field strength $B_{\rm surf}$, magnetic-field strength at the light cylinder $B_{\rm LC}$, and characteristic age $\tau_{\rm c}$.  These are given in Table~\ref{t:timing} for $d = 840$\,pc.  Constant lines of $\dot{E}$, $B_{\rm surf}$, and $\tau_{\rm c}$ are displayed in Fig.~\ref{f:ppdot} to show the distance-dependent ranges.

Spider pulsars in very-short-period orbits are difficult to discover with traditional radio searches.  Even though we can now fold the radio data with the exact parameters, \psr is still not visible.  There are two simple explanations for the non-detection of radio pulsations.  (1) Radio emission is blocked by material produced by the pulsar evaporating its companion.  Eclipses for large fractions of the orbit would be expected, since they have been seen for many spider pulsars \citep[see, e.g.,][]{fruchter1988,archibald2009,polzin2020}.  This is further supported by the observed extremely compact orbit and the strong \ac{IBS}.  Radio imaging observations could be used to check whether there is any continuum radio flux at the sky position of \psr, but previous experience is not encouraging.  The eclipses of a few other spider systems have been imaged at low frequencies, showing that, during the eclipse, the continuum flux from the pulsar disappears in tandem with the pulsed flux \citep{broderick2016,polzin2018}. (2) \psr is intrinsically radio-quiet, in that its radio beam does not cross the line of sight, or it has a very low luminosity.  There is one other radio-quiet \ac{MSP} known \citep{clark2018}.

The minimum average density of the companion $64\,\textrm{g}\,\textrm{cm}^{-3}$ is very high, assuming a filled Roche lobe \citep{eggleton1983}.  Using the filling factor from optical modeling, the average companion density $73\,\textrm{g}\,\textrm{cm}^{-3}$ is even higher.  The high density and the compact orbit suggest that the companion may be a helium white-dwarf remnant, and that the system may have evolved from an ultracompact X-ray binary \citep{sengar2017,kaplan2018}.  In addition, simulations predict evolved ultracompact X-ray binaries to have orbital periods of around $70-80$\, min \citep{haaften2012}, consistent with the $75$\,min orbital period from \psr.  Future analysis of optical spectroscopic data may give additional insight into the evolution and composition of the companion.

The discovery of \psr is the result of a multiwavelength campaign.  The pulsar-like gamma-ray spectrum, and the non-detection of radio pulsations, motivated the search for a visible companion.  This was subsequently discovered in optical and X-ray observations.  Further optical observations provided constraints on the orbital parameters that were precise enough to enable a successful gamma-ray pulsation search.

%%%%%%%%%%%%%%%%%%%%%%%%%%%%%%%%%%%%%%%%%%%
\acknowledgments{}

% Einstein@Home
We are deeply grateful to the thousands of volunteers who donated their computing time to Einstein@Home, and to those whose computers first detected \psr: Yi\nobreakdash-Sheng Wu of Taoyuan, Taiwan; and Daniel Scott of Ankeny, Iowa, USA.

% people
This work was supported by the Max-Planck-Gesellschaft~(MPG), by the Deutsche Forschungsgemeinschaft~(DFG) through an Emmy Noether Research grant, No. PL~710/1-1 (PI: Holger~J.~Pletsch) and by National Science Foundation grants 1104902 and 1816904.
L.N. was supported by an STSM Grant from COST Action CA16214.
C.J.C. and R.P.B. acknowledge support from the ERC under the European Union's Horizon 2020 research and innovation program (grant agreement No. 715051; Spiders).
V.S.D. and ULTRACAM are supported by the STFC. R.W.R. and D.K. were supported in part by NASA grant 80NSSC17K0024.
S.M.R. is a CIFAR Fellow and is supported by the NSF Physics Frontiers Center award 1430284 and the NASA Fermi GO Award NNX16AR55G.
Fermi research at NRL is funded by NASA.
J.W.T.H. is an NWO Vici fellow.

% optical
The ULTRACAM photometry was obtained as part of program WHT/2015A/35. The William Herschel Telescope is operated on the island of La Palma by the Isaac Newton Group of Telescopes in the Spanish Observatorio del Roque de los Muchachos of the Instituto de Astrof\'{i}sica de Canarias.
Based on observations made with the Isaac Newton Telescope (program I17BN005) operated on the island of La Palma by the Isaac Newton Group of Telescopes in the Spanish Observatorio del Roque de los Muchachos of the Instituto de Astrof\'isica de Canarias.
This paper makes use of data obtained from the Isaac Newton Group of Telescopes Archive which is maintained as part of the CASU Astronomical Data Centre at the Institute of Astronomy, Cambridge.

% radio
We acknowledge support of the Department of Atomic Energy, Government of India, under project No. 12-R\&D-TFR-5.02-0700 for the GMRT observations. The GMRT is run by the National Centre for Radio Astrophysics of the Tata Institute of Fundamental Research, India.
The Nan\c{c}ay Radio Observatory is operated by the Paris Observatory, associated with the French Centre National de la Recherche Scientifique (CNRS). We acknowledge financial support from the ``Programme National Hautes Energies'' (PNHE) of CNRS/INSU, France.
This Letter is based (in part) on data obtained with the International LOFAR Telescope (ILT) under project code LC7\_018. LOFAR \citep{haarlem2013} is the Low Frequency Array designed and constructed by ASTRON.
The National Radio Astronomy Observatory is a facility of the National Science Foundation operated under cooperative agreement by Associated Universities, Inc.  The Green Bank Observatory is a facility of the National Science Foundation operated under cooperative agreement by Associated Universities, Inc.
FAST is a Chinese national mega-science facility, built and operated by  NAOC.
Partly based on observations with the $100$\,m telescope of the MPIfR (Max-Planck-Institut f\"ur Radioastronomie) at Effelsberg.

% gamma rays
The \textit{Fermi}-LAT Collaboration acknowledges generous ongoing support
from a number of agencies and institutes that have supported both the
development and the operation of the LAT as well as scientific data analysis.
These include the National Aeronautics and Space Administration and the
Department of Energy in the United States, the Commissariat \`a l'Energie Atomique
and the Centre National de la Recherche Scientifique/Institut National de Physique
Nucl\'eaire et de Physique des Particules in France, the Agenzia Spaziale Italiana
and the Istituto Nazionale di Fisica Nucleare in Italy, the Ministry of Education,
Culture, Sports, Science and Technology (MEXT), High Energy Accelerator Research
Organization (KEK) and Japan Aerospace Exploration Agency (JAXA) in Japan, and
the K.~A.~Wallenberg Foundation, the Swedish Research Council, and the
Swedish National Space Board in Sweden.

Additional support for science analysis during the operations phase is gratefully acknowledged from the Istituto Nazionale di Astrofisica in Italy and the Centre National d'\'Etudes Spatiales in France. This work performed in part under DOE Contract DE-AC02-76SF00515.

% gravitational waves
The authors thank the LIGO Scientific Collaboration for access to the data and gratefully acknowledge the support of the United States National Science Foundation (NSF) for the construction and operation of the LIGO Laboratory and Advanced LIGO as well as the Science and Technology Facilities Council (STFC) of the United Kingdom, and the Max-Planck-Society (MPS) for support of the construction of Advanced LIGO. Additional support for Advanced LIGO was provided by the Australian Research Council.
This research has made use of data, software, and/or web tools obtained from the LIGO Open Science Center (\url{https://losc.ligo.org}), a service of LIGO Laboratory, the LIGO Scientific Collaboration and the Virgo Collaboration, to which the authors have also contributed. LIGO is funded by the U.S. National Science Foundation. Virgo is funded by the French Centre National de Recherche Scientifique (CNRS), the Italian Istituto Nazionale della Fisica Nucleare (INFN), and the Dutch Nikhef, with contributions by Polish and Hungarian institutes.

%\facilities{}

\software{\textit{Fermi} Science Tools, \texttt{MultiNest} \citep{MultiNest}, ULTRACAM software pipelines, \texttt{Icarus} \citep{Breton2012+Icarus}, \texttt{psrqpy} \citep{manchester2005,pitkin2018}, \texttt{Astropy} \citep{astropy2013,astropy2018}, \texttt{matplotlib} \citep{matplotlib2007}, \texttt{NumPy} \citep{numpy2006,numpy2011}, \texttt{GalPot} \citep{mcmillan2017}, \texttt{Lightkurve} \citep{lightkurve2018}, \texttt{PRESTO} \citep{ransom2002}, LALSuite \citep{lalsuite}}

%%%%%%%%%%%%%%%%%%%%%%%%%%%%%%%%%%%%%%%%%%%
\bibliographystyle{aasjournal}

\bibliography{library}

\end{document}